\documentclass[twocolumn]{revtex4}

\tighten

\include{epsf}

\begin{document}

\title{\textsc{A Monolithic Time Stretcher for Precision Time Recording} }

\author{ Gary~S.~Varner and Larry L. Ruckman}
\affiliation{Department of Physics \& Astronomy\\
University of Hawaii at Manoa \\
2505 Correa Road  Honolulu, HI 96822  USA}

\pagestyle{headings}

\begin{abstract}
Identifying light mesons which contain only up/down quarks (pions)
from those containing a strange quark (kaons) over the typical meter
length scales of a particle physics detector requires instrumentation
capable of measuring flight times with a resolution on the order of
20ps.  In the last few years a large number of inexpensive,
multi-channel Time-to-Digital Converter (TDC) chips have become available.
These devices typically have timing resolution performance in the
hundreds of ps regime.  A technique is presented that is a monolithic
version of ``time stretcher'' solution adopted for the Belle
Time-Of-Flight system to address this gap between resolution need and
intrinsic multi-hit TDC performance.
\end{abstract}

\maketitle

\section{Background}

\begin{figure}[!htb]
\epsfxsize=3.2in
\epsfbox{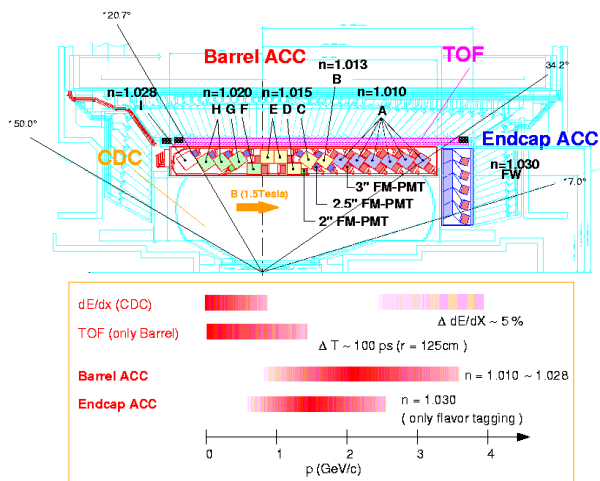}
\caption[single column]{\it Depiction of the composite detector
configuration used by Belle for particle identification.
\label{Belle_PID}}
\end{figure}

Particle identification in the Belle experiment is based upon a
composite system of subdetectors, as illustrated in
Fig.~\ref{Belle_PID}.  This hybrid system consists of ionization
loss measurements (dE/dx) in the Central Drift Chamber (CDC), Cherenkov
light emission measurement in the barrel and endcap Aerogel Chernkov
Counters (ACC), and flight time measurement in the Time Of Flight
(TOF) system.  As indicated in the lower section of this figure, the
three systems work together to cover the momentum range of interest.

Of these recording systems, the TOF system makes the most severe
demands on time resolution.  Indeed, given the 2ns spacing between RF
buckets (and possible collisions), it is not known at recording time
to which collision a given particle interaction in the TOF system
corresponds.

Precision time recording in a very high rate environment requires an
encoding scheme capable of continuous recording, with a minimum of
deadtime per logged event.  At the time of the construction of the
Belle experiment~\cite{belle} at the KEK B-factory~\cite{kekb}, a
decision was made to unify the entire detector readout (except for the
silicon vertex detector) on the LeCroy 1877 Multi-hit TDC Module.
This Fastbus module is based upon the MTD132A ASIC~\cite{mtd132a},
which has a 0.5ns resolution encoding, comparable to a number of
similar devices~\cite{MTDC1,MTDC2,MTDC3}.  Given the limited manpower
for DAQ system development and maintenance, this proved to be a wise
choice.  The intrinsic time resolution was quite adequate for
recording the timing information from the CDC, as well as the
amplitude information (through use of a charge-to-time converter) for
the CDC and ACC.

\begin{figure}[htbp]
\epsfxsize=3.2in
\epsfbox{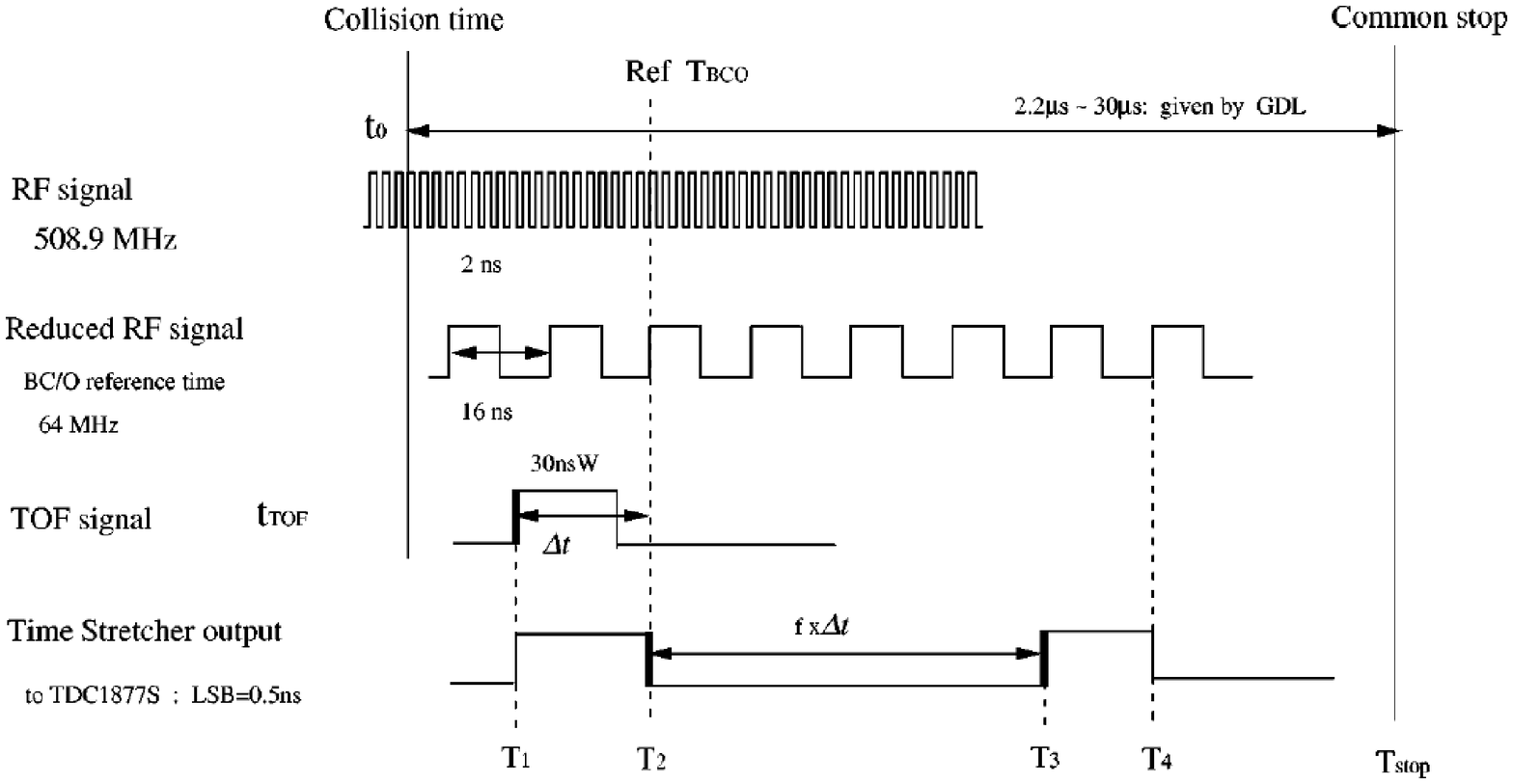}
\caption[single column]{\it Timing diagram illustrating the operating
principle of the Time Stretcher circuit, as explained in the text.
\label{TS}}
\end{figure}

The challenge then was to be able to record PMT hits with 20ps
resolution, using a multi-hit TDC having 500ps least count, and for
collisions potentially separated by only 2ns.  This latter constraint
meant that traditional techniques using a common start or stop could
not be applied, since the bunch collision of interest was not known at
the time at which the hits need to be recorded.  Moreover, in order to
avoid incurring additional error due to comparing a separate fiducial
time, it is desirable to directly reference all time measurements to
the accelerator RF clock.  The solution adopted was a so-called Time
Stretcher circuit, developed by one of the authors in conjunction with
the LeCroy Corporation~\cite{rd100}.  This work built upon valuable
lessons learned in developing a similar recording system for the
Particle Identification Detector system of the CPLEAR
experiment~\cite{cplear}.  The principle of operation is seen in
Fig.~\ref{TS}.  Hits are time-dilated with respect to the accelerator
clock and recorded at coarser resolution, but in direct proportion to
the stretch factor employed.  Statistically, by also logging the raw
hits, this stretch factor can be determined from the data.

As seen in Fig.~\ref{TS}, four timing edges are recorded for each hit
signal.  The leading edge corresponds to the actual output time of the
discriminator.  This rising edge is paired with a falling edge,
corresponding to the 2nd accelerator reference clock (RF clock divided
by 16) occuring after the initial hit timing.  The interval of
interest is then bounded to be between about 16-32 ns.  With a TDC
least count of 0.5ns, a factor of twenty time expansion is needed --
the stretch factor.  In the figure the third edge corresponds to the
time-expanded version of the interval between the rising and falling
edges.  A benefit of this technique is that it provides
self-calibration.  By recording a large number of events, the stretch
factor can be extracted from the data itself since the raw and
expanded signals are recorded.  A 4th edge is provided, two clock
rising edges after the 3rd edge, to provide a return to known state before
next pulse.  An obvious drawback in this scheme is that the deadtime
for each hit will be something like 320 - 640ns, as will be discussed
later.

\begin{figure}[htbp]
\epsfxsize=3.2in
\epsfbox{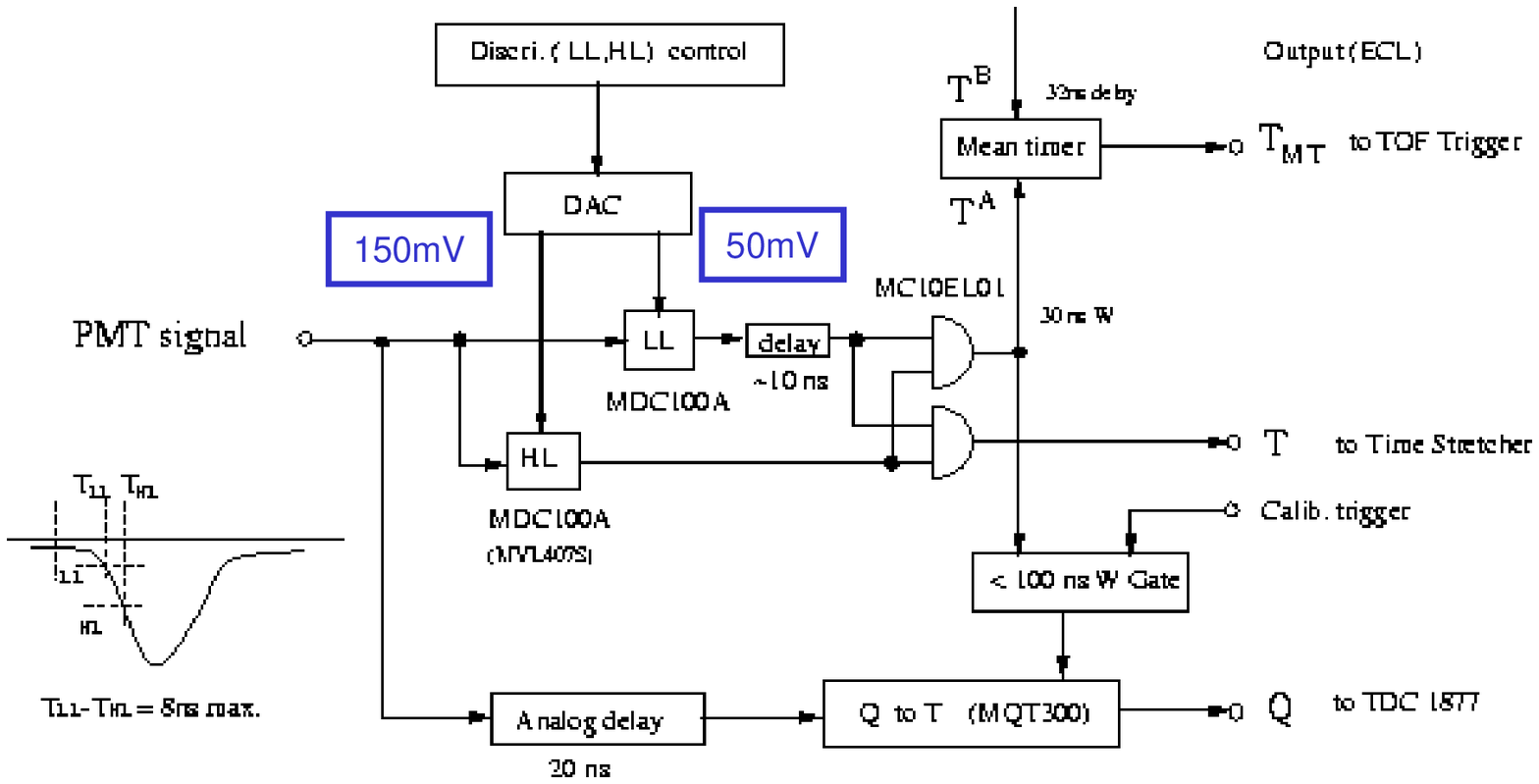}
\caption[single column]{\it Time Of Flight Front-End Electronics
readout flow.  Precision timing is performed with a coarse, multi-hit
TDC (LeCroy 1877) by means of a time-stetcher circuit.
\label{TOFFEE}}
\end{figure}

In more detail, the signal chain of the current Belle TOF
electronics~\cite{tof}, is sketched in Fig.~\ref{TOFFEE}.  

High and low level discriminators are used, with the high-level used
to reject background photon hits in the TOF and a low-level threshold
used to provide the best possible leading edge timing.  The charge of
triggered events is also recorded with a charge to time (Q-to-T) ASIC,
which is recorded with the same common TDC module.  Charge recording
is needed to correct for amplitude dependent timing effects in the
discriminator itself.

\section{Super B Factory}

The TOF readout system has worked well for almost a decade.  Increased
luminosity (already 60\% over design) has lead to much higher single
channel rates than had been specified in the design.  From the
beginning, the maximum design specification was 70kHz of single
particle interaction rate for each channel.  At this rate the expected
inefficiency would be a few percent, comparable to the geometric
inefficiency (due to cracks between scintillators).

\begin{figure}[hbtp]
\begin{center}
\epsfxsize=3.2in
\epsfbox{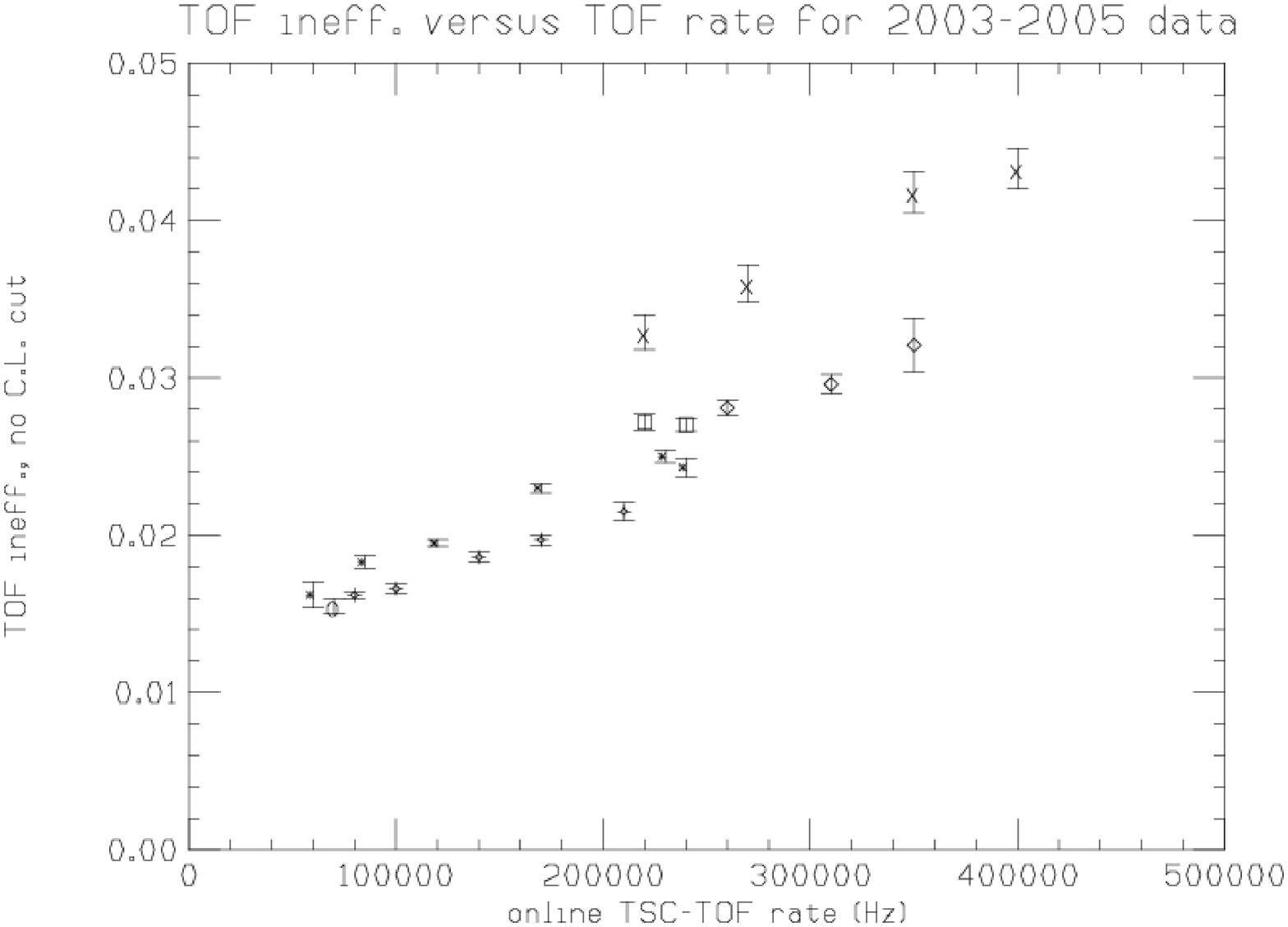}
\caption[single column]{\it Composite TOF inefficiency for the last 3
years of running at Belle.  Inefficiency grows with higher TOF singles
rates, which have increased with increased luminosity, well beyond the
70kHz design specification.
\label{occ}}
\end{center}
\end{figure}

Already the world's highest luminosity collider, the KEKB
accelerator~\cite{kekb} can now produce in excess of one million B
meson pairs per day.  Upgrade plans call for increasing this
luminosity by a factor of 30-50, providing huge data samples of 3rd
generation quark and lepton decays.  Precise interrogation of Standard Model
predictions will be possible, if a clean operating environment can be
maintained.  Extrapolation of current occupancies to this higher
luminosity mandates an upgrade of the readout electronics.  The
current system already suffers from significant loss of efficiency
with higher background rates, as may be seen in Fig.~\ref{occ}.

\section{Particle Identification Improvement}

\begin{figure*}[!htb]
\epsfxsize=5.5in
\epsfbox{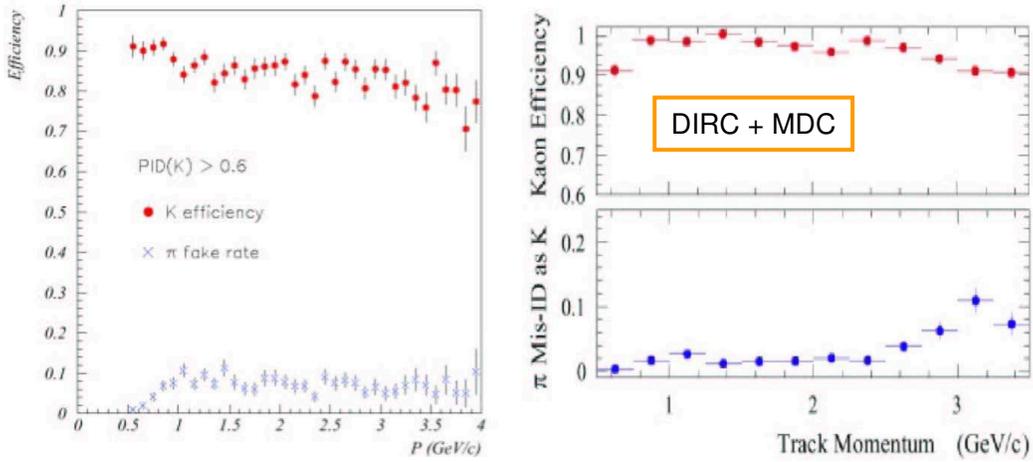}
\caption[single column]{\it A direct comparison of PID technologies
for the B-factory detectors.  On the left, the performance of the
Belle hybrid TOF/ACC system; on the right, a similar plot for the
BaBar DIRC system.  In both cases at lower momentum the K/$\pi $
separation is enhanced through the use of drift-chamber dE/dx
information.  As may be seen, the overall fake rate is lower and
efficiency higher for the DIRC.
\label{PIDbakeoff}}
\end{figure*}

In considering an upgrade to the TOF readout electronics, it is
worthwhile to consider the needs of an upgraded PID system for Belle.
A comparative study of the Belle system, as depicted in Fig.~\ref{Belle_PID}, 
with that of BaBar~\cite{DIRC} PID system is
informative.  It is clear in Fig.~\ref{PIDbakeoff} the Direct Internally
Reflected Cherenkov (DIRC) detector of BaBar has a higher efficiency
and lower fake rate than the hybrid TOF/ACC scheme used by Belle.

Indeed, it was realized in the
construction stage of Belle that such a DIRC-type detector would have
merits, and prototypes were explored~\cite{clu}.  While these results
were very promising, the schedule risks led the collaboration to stick
with technologies in which significant time and effort had already
been invested.

Thinking about an upgrade, it is reasonable to revisit the choice of
technology.  In the intervening decade, significant progress has been
made in the development of Ring Imaging CHerenkov (RICH)
detectors~\cite{RICH1,RICH2}, as well as detectors based upon the
arrival time of the Cherenkov photons, such at the Correlated
Cherenkov Timing (CCT)~\cite{cct} and Time Of Propagation (TOP)~\cite{top}
counters.

\begin{figure*}[hbtp]
\begin{center}
\epsfxsize=5.5in
\epsfbox{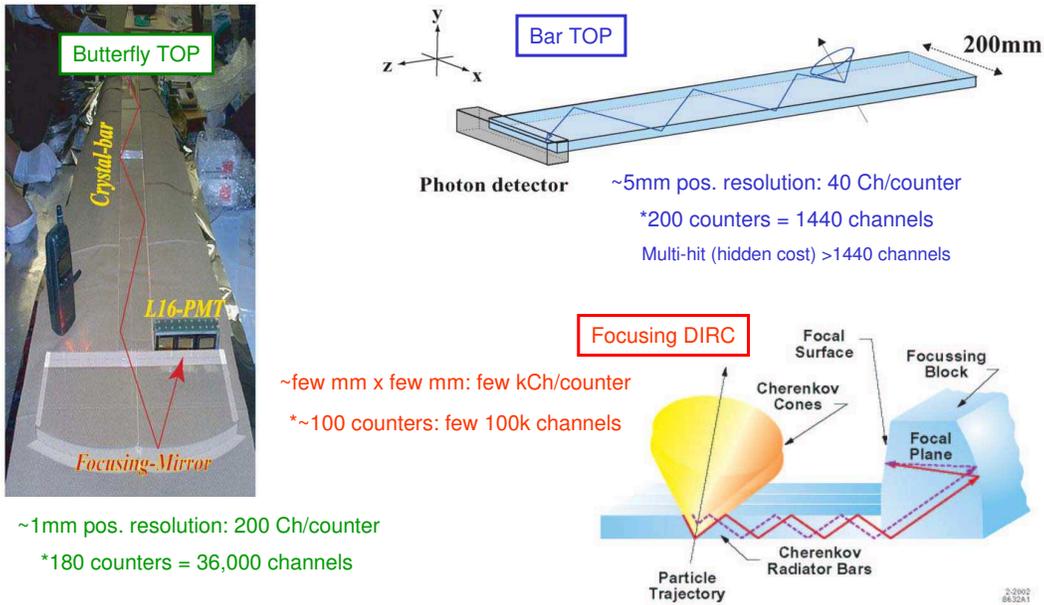}
\caption[single column]{\it Concept figures of 3 of the Cherenkov ring
imaging detectors that have been considered for the Belle detector
upgrade.  While simplest, the ``Bar TOP'' (Time-Of-Propagation)
detector has been ruled out due to inadequate performance.  Of the
remaining two, the number of instrumented readout channels will depend
upon the photodetector chosen, though will likely require many tens of
thousands of readout channels, dictating a monolithic approach.
\label{PIDconcept}}
\end{center}
\end{figure*}

Because of the great cost encumbered in the procurement and
construction of the CsI crystal calorimeter, it is planned not to
upgrade the barrel section.  As a consequence, the volume available
for the TOF/ACC replacement detector is rather limited.  Therefore a
RICH type detector has not been pursued.  The most promising
technologies to date are those illustrated in Fig.~\ref{PIDconcept}.
The TOP concept uses timing in place of one of the projected spatial
dimensions to reconstruct the Cherenkov emission ring.  A focusing
DIRC is principally using geometry to reconstruct the Cherenkov ring
segments.  However, in this case precision timing is still very useful
for two important reasons.  First it allows for the possibility of
using timing to correct for chromatic dispersion in the emission angle
of the Cherenkov photon.  And second, fine timing allows time of
flight to be measured using the quartz radiator bar.

Therefore, in both of the viable detector options considered, a large
number of fine timing resolution recording channels are required.  In
the case of a finely segmented focusing DIRC~\cite{fDIRC} option, the
number of readout channels could be comparable to that of the current
silicon vertex detector.  Clearly if such a detector is to be viable,
significant integration of the readout electronics will be essential.

Not shown is a proposal for a multi-segmented TOF detector consisting
of short scintillator bars.  While this option remains viable (and the
electronics presented would work well with such a system), the PID
performance degradation of such a system is probably unacceptable.  Of
the choices listed, the most attractive in terms of performance is a
focusing DIRC detector, if the issues of the photodetector and readout
can be addressed.

Either as an upgrade of only the readout electronics or as a prototype
for a higher channel count PID detector, it is worth considering improvements 
to the existing readout.



\section{The Monolithic Time Stretcher}

The Time Stretcher technique has worked very well and Belle has been
able to maintain approximately 100ps resolution performance with the
TOF system.  A slow degradation with time is consistent with loss of
light output.  Detailed Monte-Carlo simulation~\cite{jiwoo} has been
able to reproduce much of the performance of the TOF system and the
degradation is consistent with light loss due to crazing of the
scintillator surface.  A larger concern is the significant degradation
of TOF system performance due to high hit rates.  While the multi-hit
TDC is capable of keeping up with high rates (though the limited
number of recorded edges (16) also leads to inefficiency), by its very
nature, the Time Stretcher output can not be significantly reduced.
Recently, the clock speed was doubled, to help reduce this effect.
Nevertheless, at ever higher hit rates, the deadtime leads to ever
increasing inefficiency.

\begin{figure*}[ht]
\begin{center}
\epsfxsize=5.8in
\epsfbox{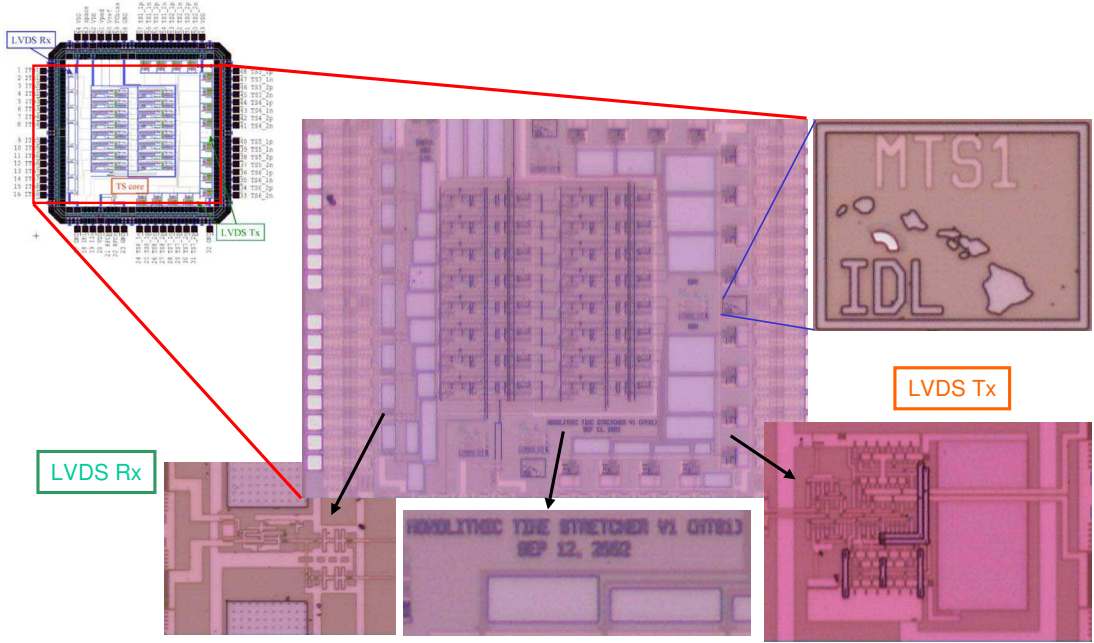}
\caption[single column]{\it Floorplan drawing and die photographs of
the Monolithic Time Strecther version 1 (MTS1) ASIC, an 8 channel
device.  All inputs and outputs are LVDS to reduce cross-talk.
\label{MTS1}}
\end{center}
\end{figure*}

A logical solution to this problem is to introduce a device which has
buffering.  Also, while taking the effort to reduce the deadtime, it
makes sense to consider a much more compact form-factor.  This was
done with the thought toward moving to a larger number of readout
channels in a future Belle PID upgrade~\cite{shuzenji}, as mentioned
earlier.  One proposed solution is the Monolithic Time Stretcher
(MTS) chip, a prototype of which is shown in Fig.~\ref{MTS1}.

The fundamental logic of the device is identical to that currently in
use with two major changes:

\begin{enumerate}
\item High density
\item Multi-hit
\end{enumerate}

High density is achieved by replacing discrete Emitter-Coupled Logic
components on daughter cards with a full custom integrated circuit.
This higher integration permits having multiple time stretcher
channels for each input.  By toggling to a secondary output channel,
the deadtime can be significantly reduced.  Once a hit is
processed in one output channel, the next is armed to process a subsequent
hit.

\begin{figure*}[hb]
\epsfxsize=5.8in
\epsfbox{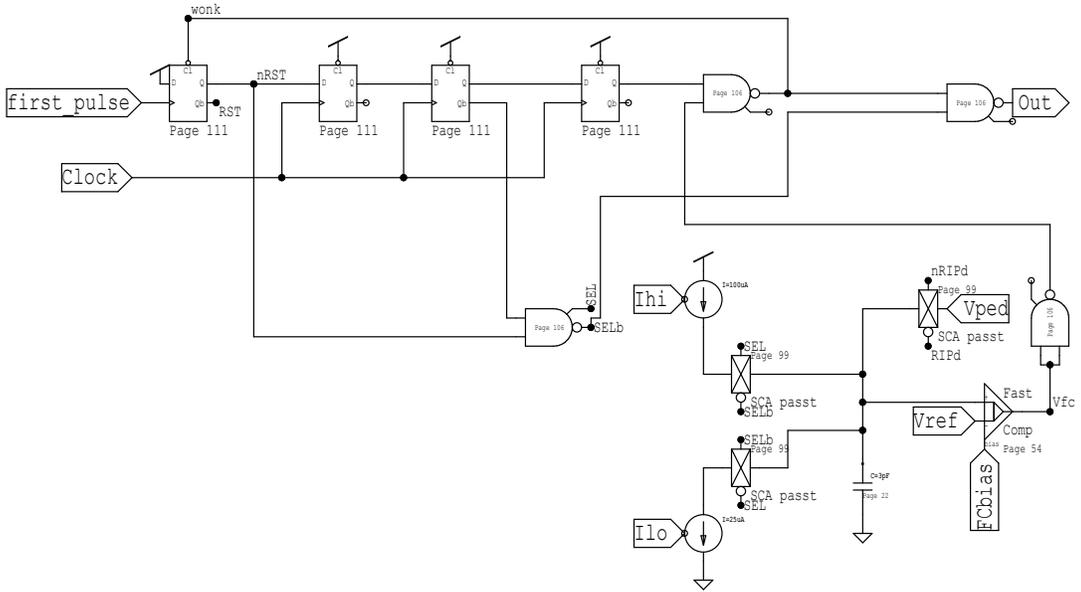}
\caption[single column]{\it Schematic of the basic clocked time stretcher circuit.
\label{TS_CKT}}
\end{figure*}

In Fig.~\ref{MTS1} the 8 channel repeating structure of each time
stretcher circuit is clearly seen in the die photograph.  The basics
of the time-stretcher circuit are visible in Fig.~\ref{TS_CKT}.  A
one-shot circuit at the upper left leads to an immediate output
signal, as well as starts charging current $I_{\rm hi}$.  Pipelining
of the hit signal continues for two clock cycles after which current
$I_{\rm hi}$ is switched off and discharge current $I_{\rm lo}$ is
switched on.  A comparator monitors the voltage induced on the storage
capacitor due to charging and discharging, providing an output signal
to indicate the stretched time when the voltage is discharged.


\clearpage

The stretch factor is given by the ratio of the two currents: $SF =
I_{\rm hi} / I_{\rm lo} $.  Each input channel of the MTS1 has two
time stretcher circuits, the second corresponding to the secondary
output when the primary channel is active.  Each output is recorded by
a separate TDC channel.  With this configuration at 10\% deadtime for
a single channel of time stretcher can be reduced to 1\%.  As the 
the incremental cost of additional TDC channels is rather low, it
is possible consider additional buffering depths, which would reduce
the deadtime by the $dT^N$, where N is the buffer depth, though that
was not explored beyond a depth of two in this device.

Reduction of cross-talk and Electro-Magnetic Interference is enhanced
by the use of Low Voltage Differential Signalling (LVDS) \cite{LVDS}.
MTS1 is fabricated in the Taiwan Semiconductor Manufacturing
Corporation $0.35\mu $m CMOS process.  


\subsection{Form Factor Reduction}

\begin{figure}[htbp]
\epsfxsize=3.2in
\epsfbox{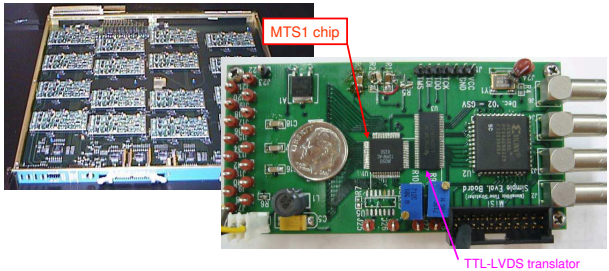}
\caption[single column]{\it A form-factor comparison between the
current Fastbus-sized, 16-channel Time Stretcher and the MTS1 chip on
a test board.  The test board occupies almost the same space as a
single daughtercard channel on the TS motherboard, and has the same
number of channels of time-stretching as the whole module.
\label{MTS_comp}}
\end{figure}

When considering a photodetector with a large number of channels, the
form factor of this device is very attractive, as shown for comparison
in Fig.~\ref{MTS_comp}, a substantial reduction in size has been
achieved.  On the left is a 16-channel Fastbus-sized Time Stretcher
card used currently in the Belle experiment.  Inset is a test board
with one of the MTS1 packaged devices for comparison, where a dime has
been placed on the board for scale.  

With this level of integration it becomes feasible to consider
integration of the time stretcher and TDC electronics on detector, as
is being done for detector subsystems in the LHC experiments.  

\subsection{Test Results}

\begin{figure}[htbp]
\epsfxsize=3.2in
\epsfbox{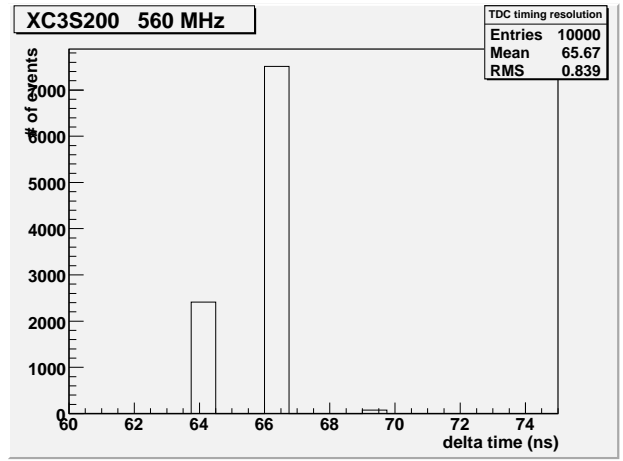}
\caption{\it Timing resolution obtained for the FPGA-based TDC used in the MTS1 evaluatiton.
\label{Timing_resol} }
\end{figure}

In order to test the performance of the MTS1, a multi-hit TDC should be
used.  As a demonstration of the power of this time stretching
technique, an Field Programmable Gate Array (FPGA) can be used as this TDC~\cite{GSV_JINST}, where the
results from a simple Gray-code counter implementation of the hit time
recording may be seen in Fig.~\ref{Timing_resol}.  The RMS of the
distribution is about 840ps for the Xilinx Spartan-3 device used.
This resolution could be improved by use of a faster FPGA, though is
sufficient to obtain the test results shown below.

Indeed, it is worth noting that this combined Time-Stetcher + FPGA
technique is very powerful for two important reasons:

\begin{enumerate}
\item low-cost, high-density TDC implementation
\item deep and flexible hit buffering and trigger matching logic
\end{enumerate}

\begin{figure}[htbp]
\epsfxsize=3.2in
\epsfbox{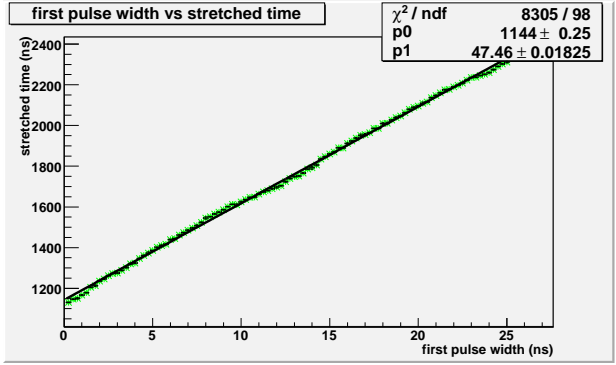}
\caption[single column]{\it Scan of stretched times versus input
reference time, within a single stretch clock cycle.  In this case a
stretch factor of about 47.5 was used.
\label{MTS1_linear}}
\end{figure}

A test sweep of the MTS1 input is shown in Fig.~\ref{MTS1_linear},
where it should be noted that due to the encoding scheme it is only
meaningful to scan within a time expansion clock cycle period.  A scan
of expansion ratios was performed and the best results were obtained
for stretch factors of 40-50.

\begin{figure}[htbp]
\epsfxsize=3.2in
\epsfbox{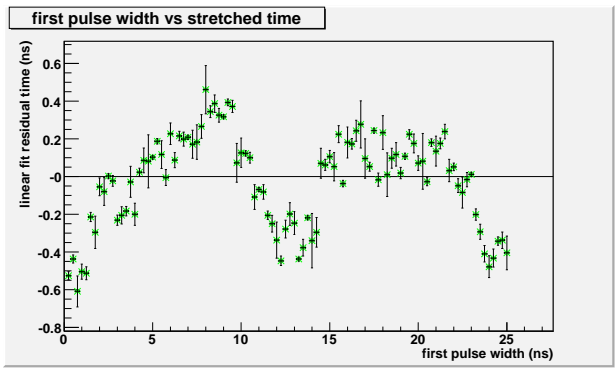}
\caption[single column]{\it Residual distribution after a linear fit
and application of the time stretch factor.  An effect of expansion
clock is clearly seen.
\label{MTS1_resid}}
\end{figure}

As can be seen, there is some non-linearity in the expanded time.
This is more clearly seen when a plot of the residual distribution is
made by subtracting off the linear fit, as shown in
Fig.~\ref{MTS1_resid}.  A periodic structure is seen, roughly
consistent with the expansion clock period, if the negative timing
dips are correlated to transition edges.

\subsubsection{Timing Resolution}

\begin{figure}[htbp]
\epsfxsize=3.2in
\epsfbox{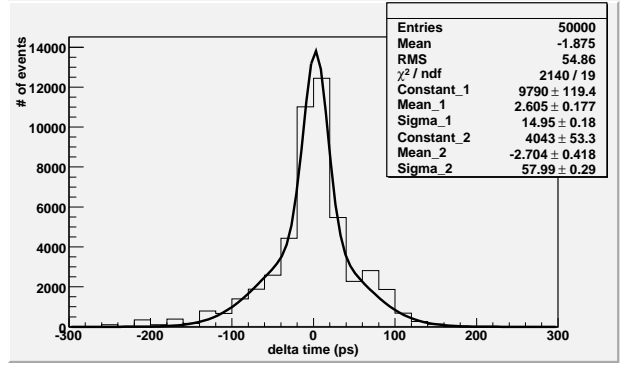}
\caption[single column]{\it Timing resolution of the MTS1 + FPGA TDC when a
non-linearity correction is applied.  Non-gaussian tails are due to
regions with larger jitter due to coupling of the reference clock into
the ramping circuitry.
\label{MTS1_resol}}
\end{figure}

As with the HPTDC device~\cite{HPTDC} developed at CERN for the ALICE
detector, a fine calibration is needed to obtain a precision
comparable to the current Belle system.  Applying such a calibration,
determined in a separate data set, significantly improved linearity
and residuals are obtained.  The subsequent results are histogrammed
in Fig.~\ref{MTS1_resol}.

As can be seen, the timing resolution fits well to a double Gaussian,
with a narrow sigma less than 20ps, which is comparable to (and
actually slightly better than) the existing Belle system.  This result is
consistent with the expectation from the FPGA TDC used, where

\begin{equation}
\sigma_{\rm TS} = {840{\rm ps} \over {\rm Stretch Factor}} \simeq
{840{\rm ps} \over 47.5 } \approx 17{\rm ps}
\end{equation}

and the measured sigma is about 15ps.  It is possible that a finer
resolution FPGA TDC would allow for an even more precise timing determination.

In practice the systematic effects of the upstream discriminator and
its amplitude dependent threshold crossing (and comparator overdrive)
dependence make any further improvements difficult.  Nevertheless it
is an interesting question for future exploration.  This timing
resolution is comparable to that obtained with the HPTDC after careful
non-linearity calibration.

The broader Gaussian distribution and significant non-gaussian tails
are correlated with expansion clock feedthrough to the ramping
circuit.  This could be improved in a future version with better
layout isolation.  The $0.35\mu$m process used only had 3 metal
routing layers available and migration to a finer feature size process
would allow for dedicated shields and better power routing.

\subsubsection{Multi-hit Buffering}

In order to reduce deadtime a second time-stretcher circuit, with a
separate output, is provided for each input channel.  This second
circuit becomes armed when the primary stretcher circuit is running.
Use of such a scheme can significantly reduce data loss due to arrival
of subsequent hit during operation of the first stretcher circuit.  The factor
may be expressed as

\begin{equation}
F_{\rm dead} = F_{\rm single}^N
\end{equation}

where N is the number of buffer stages. For N=2, the case prototyped
here, a large existing deadtime of 20\% could be reduced to 4\%.
Moreover, this technique can be extended to an even larger number of
buffer channels, a realistic possibility when using a low cost
FPGA-based TDC.  In the case of 4 outputs, a 20\% single time
stretcher deadtime would become a completely negligible $1.6\times
10^{-3}$.

\begin{figure}[htbp]
\epsfxsize=3.2in
\epsfbox{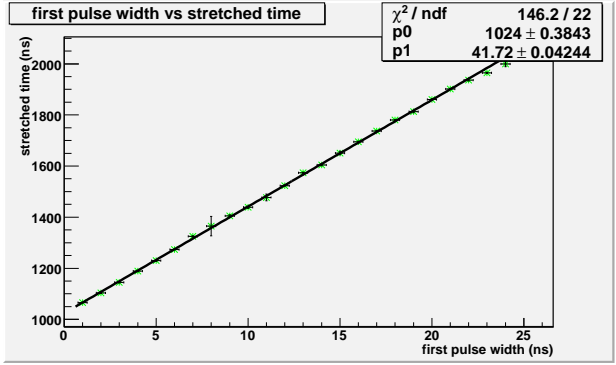}
\caption[single column]{\it Cross-check measurement of the secondary
MTS1 output channel, where the results are seen to be comparable
to the primary channel, apart from a systematically smaller stretch
factor as described in the text.
\label{MTS1_buff}}
\end{figure}

Apart from the arming circuitry, the second time stretcher channel is
identical to the primary.  Testing was performed with double-pulse
events and the result for the second channel is seen in
Fig.~\ref{MTS1_buff}.

Note that these secondary channels have a time-stretch factor that is
systematically smaller.  As the same reference currents are mirrored
in all channels, it is believed that this is due to ramp window
reduction due to latency in the arming logic.

\subsubsection{Cross-talk}

\begin{figure}[htbp]
\epsfxsize=3.2in
\epsfbox{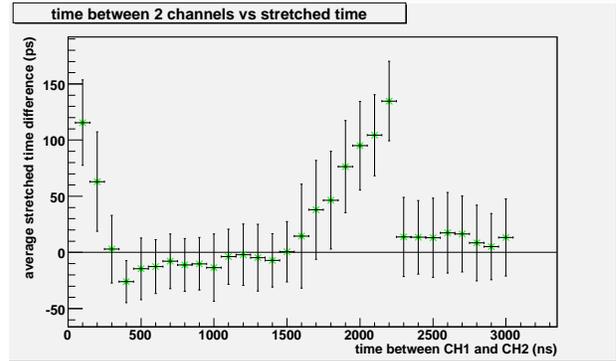}
\caption[single column]{\it Timing shift due to adjacent channel
crosstalk.  As expected, impact is most sensitive during the initial
current ramping and near stretched time threshold crossing.
\label{NTS32_crosstalk}}
\end{figure}

An important check of performance of the MTS1 is the impact of time
stretcher operation on one channel while another is operating.  This
has been performed in Fig.~\ref{NTS32_crosstalk}, where the timing of
the first channel is fixed and the timing relation of the signal in
channel 2 is varied.

The impact of operation of this second channel is clear during the
ramping portion of the readout cycle, as well as the threshold
crossing at the end of the ramping interval.  While this effect can be
calibrated out to some extent, just like effects of the clock feedthrough,
this perturbation to the circuit would be better mitigated through
better isolation in the IC layout.


\section{Future Prospects}

An improved layout paired with future, higher clock frequency FPGAs
could open the possibility of very dense channel count, sub-10ps
resolution TDC recording.

For many applications the HPTDC is perfectly suitable and gives
comparable time resolution to the MTS1 + FPGA TDC.  In both cases a
non-linearity correction is required to obtain this resolution.
However the time encoding itself is only part of the issue for
obtaining excellent timing resolution from a detector output.
Correction for time slew in the discriminator threshold crossing is
critical.  Moreover the addition of many channels of high-speed
discriminator inside a detector is a noise and power concern.
Compact, high-speed waveform recording~\cite{BLAB1} may be a promising
next evolutionary step in the readout of precision timing detectors.

\section{Acknowledgements}

This work was supported by the US-Japan Foundation and the Department of
Energy Advanced Detector Research Award Number DE-FG02-06ER41424.

\clearpage

\setlength{\parindent}{0ex}

\end{document}